\documentclass[letterpaper, 10pt, conference]{ieeeconf}

\IEEEoverridecommandlockouts                              

\usepackage{graphics} 
\usepackage{amsmath} 
\usepackage{amssymb}  
\usepackage{svg}
\usepackage[noadjust]{cite}
\usepackage{caption}
\usepackage{multirow}
\usepackage{multicol}
\usepackage{subfigure}
\usepackage{hyperref}
\hypersetup{
    colorlinks=true,
    linkcolor=blue,
    filecolor=magenta,      
    urlcolor=blue,
}
 
\title{\LARGE \bf Individual and Team Trust Preferences for Robotic Swarm Behaviors}
\usepackage[belowskip=-4pt,aboveskip=0pt]{caption}

\author{Elena M. Vella$^{1}$,  Daniel A. Williams$^{2}$, Airlie Chapman$^{1}$, Chris Manzie$^{2}$  
\thanks{$^{1}$Elena M. Vella and Airlie Chapman are with the Department of Mechatronic Engineering,
        The University of Melbourne, 3052 Parkville, Australia
        {\tt\small evella@student.unimelb.edu.au}, {\tt\small airlie.chapman@unimelb.edu.au}.}%
\thanks{$^{2}$Daniel A. Williams and Chris Manzie are with the Department of Electrical Engineering, The University of Melbourne, 3052 Parkville, Australia
        {\tt\small williamsd4@student.unimelb.edu.au}, {\tt\small manziec@unimelb.edu.au}.}%
        }
        
\begin{document}

\maketitle
\thispagestyle{empty}
\pagestyle{empty}

\begin{abstract}
Trust between humans and multi-agent robotic swarms may be analyzed using human preferences. 
These preferences are expressed by an individual as a sequence of ordered comparisons between pairs of swarm behaviors. 
An individual's preference graph can be formed from this sequence.
In addition, swarm behaviors may be mapped to a feature vector space. 
We formulate a linear optimization problem to locate a \textit{trusted behavior} in the feature space. 
Extending to human teams, we define a novel \textit{distinctiveness} metric using a sparse optimization formulation to cluster similar individuals from a collection of individuals' labeled pairwise preferences.
The case of anonymized unlabeled pairwise preferences is also examined to find the average trusted behavior and minimum covariance bound, providing insights into \textit{group cohesion}.
A user study was conducted, with results suggesting that individuals with similar trust profiles can be clustered to facilitate human-swarm teaming.

\end{abstract}
\section{Introduction}\label{intro}

As teams of robots venture further into home and work settings, interactions between humans and robotic systems will become more common. 
An important element of these interactions is the notion of human trust, an intuitive belief that the object of trust will help to achieve goals even in uncertain situations \cite{lee_trust_2004}. 
Excessive trust in a system can lead to over-reliance, with potentially negative implications for performance and safety. Conversely, humans with low trust may avoid interacting with the system altogether. 
It is vital to understand how the design and operation of robotic systems influences human trust perceptions, in order to facilitate more efficient and successful interactions. 
We therefore focus on learning and predicting human trust in robotic swarms. 

Much attention has focused on trust in human-automation interaction, initially in human interactions with industrial machines \cite{lee_trust_2004} but later expanding to other settings \cite{lyons_human-autonomy_2021}. Trust in human-robot interaction is an active area of research \cite{schaefer_measuring_2016}, stemming from scenarios that require a delegation of autonomy (e.g. collaborative lifting, search and rescue). 
This is particularly salient in human interactions with multi-robot swarms (HSI), as it is difficult for a human to interact with each swarm agent simultaneously. 
Delegation of autonomy can only happen when the human trusts the swarm sufficiently (regarding proper task execution and safety, among other factors \cite{schaefer_measuring_2016}). 
Conventional approaches to measuring human trust in swarms use a scalar value \cite{nam_models_2020} or multi-dimensional quantity \cite{schaefer_measuring_2016} to represent trust, however the nebulous, idiosyncratic nature of trust may complicate attempts to compare values between individuals and across time. 
In addition, trust measures may focus on certain characteristics and omit others, prescribing a model of trust that may not suit all participants. 
For greater comparability when analyzing trust, we may instead focus on human preferences and use preference learning techniques \cite{furnkranz2010preference}.
Inspired by the algorithm presented in \cite{kingston_comparing_2009}, we may use preferential reasoning to understand the influence of swarm behaviors on a human observer's trust level. 
We can further extend the concept of individual trust to groups \cite{akash_dynamic_2017}, thus providing a way of understanding how to team individuals based on similar trust profiles and of analyzing conflicting preferences' impact on a team's overall trust dynamic.

We extend concepts developed in \cite{furnkranz2010preference, kingston_comparing_2009, akash_dynamic_2017} to population based measurements of trust. 
Our contributions include a \textit{distinctiveness} metric describing how an individual's trust towards a swarm differ from others in a population. 
We determine this metric this by analyzing differences in individuals' trust preferences as perturbations from a common reference, and quantifying the respective divergence for each individual.
By selecting those with distinctiveness below a threshold, we may cluster individuals with similar trust preferences.
We also consider the concept of \textit{group cohesion} regarding the distribution of trust preferences when preferences are anonymized and aggregated from a population.
To this end, assuming the individual's optimal trust is drawn from a normal distribution, the unlabeled preference data can yield bounds on the distribution covariance. 
These in turn can serve as a measure of group cohesion.

The remainder of this paper studies how we may analyze human trust preferences for robotic swarm behaviors and describe a group's trust preferences by synthesizing individuals' preferences.
We first define key terminology and explain the swarm behaviors considered. 
In $\S$\ref{formulation} we formulate a preference learning system and introduce \textsc{Valma} for feature vector extraction. 
We develop a trust preference model for an individual in $\S$\ref{trust_ind}, extending this to a group in \ref{trust_gr}.
Details of a user study feature in $\S$\ref{study}, and concluding remarks are made in $\S$\ref{conclusions}. 

\paragraph*{Notation}
For $x\in\mathbb{R}^{q}$, we define the 1-norm of a vector $||x||_{\mathbf{1}}=\sum_{i=1}^{n}|x_{i}|$ and the 2-norm of a vector $||x||_{\mathbf{2}}=\sum_{i=1}^{n}|x_{i}|^{2}$.
The indicator function $\mathbb{I}_{A}(\cdot)$ for a set $\mathcal{A} \in \mathbb{R}$ is defined as $\mathbb{I}_{A}(x) = 1$ if $x\in \mathcal{A}$ and otherwise equal to 0.
The identity matrix $I$ is a diagonal matrix with 1 on the dominant axis and 0 elsewhere.
A pair of elements $(x^{1}_i, x^{2}_i)$ appears uniquely in a sequence set $\mathcal{S}$, if and only if $x^{1}_i$ is preferred to $x^{2}_i$ for all possible set combinations $i = 1,...,p$.
A graph $\mathcal{G}=(V,S)$ is defined by a vertex set $V$ of cardinality $n$ and an ordered edge set $E \subseteq V \times V$ of cardinality $m$.
If an edge exists from vertex $v_{j}$ to vertex $v_{i}$ it is expressed as $(v_j,v_i)\in E$. A directed acyclic graph is a graph with no directed cycles.
We denote the standard normal cumulative density function as $\Phi(x)$. The function
evaluates the probability that the value of a random variable $Y\sim\mathcal{N}(0,1)$
is less than or equal to $x\in(-\infty,\infty)$.
Similarly, the normal cumulative density function $\Phi_{\mu,\sigma}(x)$
describes the probability that the value of the random variable $X\sim\mathcal{N}(\mu,\sigma^{2})$
is less than or equal to $x$, denoted by $p(X\leq x)$. The cumulative density functions are related by $
p(X\leq x)=\Phi_{\mu,\sigma}(x)=\Phi\left(x-\mu\right)/\sigma$; the inverse mapping is given by
\begin{equation}
\Phi^{-1}\left(p(X<x)\right)=\frac{x-\mu}{\sigma}.\label{eq:Inverse distribution mapping}
\end{equation}

\section{Problem Formulation} \label{formulation}
We seek to formulate a preference learning problem in order to study how a swarm's behavior might influence a human observer's trust and, more generally, trust for a group.

\subsection{Swarm Behaviors}\label{swarmbehaviors}

In this work we consider a swarm composed of ground-based robotic vehicles \cite{schoof_experimental_2018}. Each agent's motion can be represented using unicycle dynamics commanded by a multi-agent controller. 

\begin{figure}[thpb]
    \centering
    \subfigure[\href{https://vimeo.com/593164326}{Cyclic Pursuit}]{
        \includegraphics[width=0.32\columnwidth]{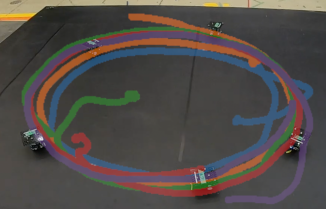}
        \label{fig:tt_cyclic}
    }
    \subfigure[\href{https://vimeo.com/593165850}{Herding}]{
        \includegraphics[width=0.32\columnwidth]{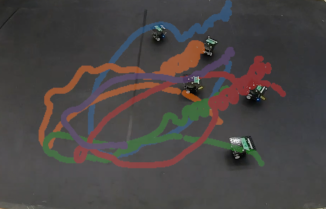}
        \label{fig:tt_herding}
    }
    \subfigure[\href{https://vimeo.com/593166047}{Follow the Leader}]{
        \includegraphics[width=0.32\columnwidth]{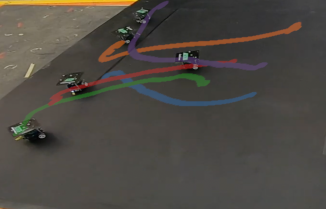}
        \label{fig:tt_lf}
    }
    \subfigure[\href{https://vimeo.com/593166861}{Square Formation}]{
        \includegraphics[width=0.32\columnwidth]{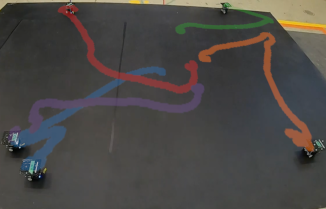}
        \label{fig:tt_square}
    }
    \subfigure[\href{https://vimeo.com/593167179}{Line Formation}]{
        \includegraphics[width=0.32\columnwidth]{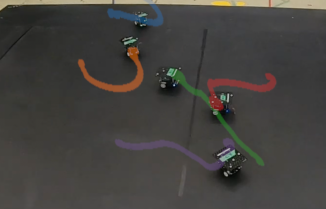}
        \label{fig:tt_line}
    }
    \caption{Swarm behavior trajectory traces.}
    \label{fig:trails}
\end{figure}

To elicit human trust responses, we have implemented five swarm behaviors using the robotic platform: \textbf{1)} \textbf{cyclic pursuit}, in which agents traverse a circle \cite{marshall_formations_2004}, \textbf{2)} \textbf{herding}, in which agents move from one location to another while maintaining collective cohesion \cite{pierson_bio-inspired_2015}, \textbf{3)} \textbf{leader following}, in which a leader moves while trailed by all other agents \cite{pierpaoli_reinforcement_2019},
\textbf{4)} \textbf{square formation}, in which agents relocate to the vertices of a square shape \cite{pierpaoli_reinforcement_2019},\textbf{ 5)} \textbf{line formation}, in which the agents relocate to form a line \cite{pierpaoli_reinforcement_2019}.
The respective trajectories are depicted in Figure \ref{fig:trails}.

\subsection{Preference Data and Preference Graph}\label{Preference graph}
Given a set of $n_{b}$ swarm behaviors $B$, we wish to collect a set of $m_{b}$ pairwise comparison preferences $\mathcal{E}_{B}=\{(v_{i}^{1},v_{i}^{2})\,|\,v_{i}^{1},v_{i}^{2}\in B,\,i\in\{1,...,m_{b}\}\}$ that answer the general question ``Comparing these two swarm behaviors, which do you trust more?", presupposing an individual's intuitive definition of trust.
For the $i^{th}$ comparison, we record the more trusted behavior as $v_{i}^{1}$ (the first element in the pair) and the less trusted behavior as $v_{i}^{2}$ (the second element).

We may visualize an individual's pairwise comparison preferences using a preference graph. 
Consistent preferences may be depicted as acyclic graphs, yielding a partial order over preferences, while inconsistent preferences generate a cyclic graph.
For the $k^{th}$ individual, the directed preference graph $\mathcal{G}_{k}=(V_{k},\mathcal{E}_{k})$ is defined by a vertex set $V_{k}\subseteq B$ containing the compared behaviors, and an ordered edge set $\mathcal{E}_{k}\subseteq V_{k}\times V_{k}\subseteq\mathcal{E}_{B}$ indicating preferences among pairs of vertices. 
Here, a directed edge $(v_{i},v_{j})\in\mathcal{E}_{k}$ from vertex $v_{i}$ to vertex $v_{j}$ indicates behavior $v_{i}$ is preferred to behavior $v_{j}$. Note there is at most one edge between each vertex in a pair, i.e. we do not consider self-contradictions. 

We assume that each preference is labeled as belonging to a given individual when constructing the individual's directed preference graph.
In a wider population for which demographic information cannot be collected, we instead collect anonymized \textit{unlabeled} trust preferences such that we cannot distinguish between individuals.
In this case we may define a population preference graph $\bar{\mathcal{G}}=(\bar{V},\bar{\mathcal{E}},W)$, with the vertex set $\bar{V}=\bigcup_{k=1}^{p}V_{k}$, the ordered set of edges $\bar{\mathcal{E}}\subseteq\mathcal{E}_{B}$, and the associated edge weight set $W$.
The edge and weight set are formed by enumerating each pairwise preference as $a_{ij}=\sum_{k=1}^{p}\mathbb{I}_{\mathcal{E}_{k}}\left((v_{i},v_{j})\right)$ for all $v_{i},v_{j}\in\overline{V}$.
The edge set $\bar{\mathcal{E}}$ contains the edge $(v_{i},v_{j})$ if $a_{ij}-a_{ji}>0$ (i.e. $v_{i}$ is more preferred to $v_{j}$) or $a_{ij}=a_{ji}\neq0$ and $i<j$ (i.e. $v_{i}$ are equally preferred). 
The associated edge weight element of $W$ is $w_{k}=a_{ij}/(a_{ij}+a_{ji})$ for the $k^{th}$ preference pair $(v_{k}^1,v_{k}^2)=(v_{i},v_{j})$; these weights capture preference variability among individuals in the population.

We interpret $\bar{\mathcal{G}}$
as proportional to the average preference of the population, with
each vertex in the graph mapped to a preferred preference. The directed
weighted edges between vertices indicate the likelihood of the preference
transition by the majority of individuals.
A partially ordered set of instance preferences can be abstracted from the preference graph and is amenable to an ordinal optimization problem \cite{ho1992ordinal}. We instead locate the preference instances within a feature vector space and optimize within this space.

\subsection{Feature Vector Extraction}
\begin{figure}[t]
\centering
\includegraphics[width=0.8\columnwidth]{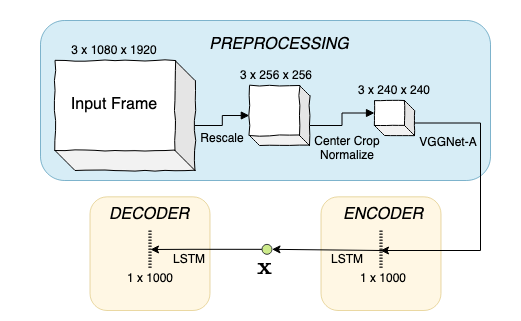}
\caption{Overview of data flows in the \textbf{V}isual \textbf{a}nd \textbf{L}ongitudinal \textbf{M}otion \textbf{A}utoencoder model architecture; note that the blocks convey the dimensionality of the signals while the arrows indicate performed transformations.} 
\label{valma}
\end{figure}

As we cannot reason about preferences towards behaviors not already in the graph, we may consider mapping each behavior to a point $x$ in a feature space $\mathcal{X}$. 
In $\S$\ref{study} we present individuals with videos of swarm behaviors. 
We seek to work directly with these stimuli, to capture information about the swarm's visual appearance and trajectory evolution encoded therein. 
Manual feature vector extraction from stimuli has been demonstrated for preference learning in \cite{kingston_metric_2014}, however doing so for videos is impractical and entails discretionary judgements regarding which features to select.
Automatic feature vector extraction processes can address these two issues by removing human discretion in feature identification.
One may consider dimensionality-reduction techniques for individual video frames (e.g. principle component analysis \cite{brunton_data-driven_2019}), however temporal dynamics between frames will be neglected.
To overcome this issue, we have adopted a similar approach to \cite{aberman_learning_2019} by developing a neural network-based variational auto-encoder (\textsc{Valma}) that extracts a feature vector for a video of swarm behavior.
As depicted in Figure \ref{valma}, the model contains a pre-processing component (extracting frame features using the \textsc{VGGNet-A} computer vision model \cite{simonyan_very_2014}) and two additional recurrent neural network components: an encoder and a decoder. 
The recurrent neural networks use an LSTM architecture \cite{hochreiter_long_1997} in order to learn temporal relationships between different data features.
Between the two components there is a data bottleneck, with the encoded input projected into a lower-dimensional latent space.\footnote{The reader is referred to \href{https://gitlab.com/williamsdaniel888/valma}{our repository} for further details about the implementation and training of the model.} 
By training the model to reproduce the encoder input at the decoder output, we can learn a mapping from the input space to a latent space in an unsupervised manner.
and thus extract a compact feature vector $x\in\mathcal{X}$ automatically. 
We describe the mapping as $h:B\to\mathcal{X}$ from the behavior set to a $q$-dimensional feature space $\mathcal{X}\subseteq\mathbb{R}^q$. 


\section{Trust for Individuals}\label{trust_ind}
To meaningfully compare individuals' preferences regarding videos of swarm behaviors, we consider pairwise comparisons (`instances') in a feature vector space. 
In the following sections we synthesize and build on \cite{kingston_comparing_2009} and \cite{kingston_metric_2014}, posing a convex optimization problem with a global extremum.

\subsection{Preference Synthesis}
The feature vectors for the trust preference pair $(v^1_i,v^2_i)\in\mathcal{E}_{B}$ are given by $x^1_i=h(v^1_i)$ and $x^1_i=h(v^1_i)$, respectively. Pairwise trust preferences imply the existence of an underlying quadratic trust function $f_{k}: \mathcal{X} \to \mathbb{R}$ such that
\begin{equation}\label{eq:costComparison}
f_{k}(x^{1}_i) \leq f_{k}(x^{2}_i) \Leftrightarrow (v^{1}_i,v^{2}_i)\in \mathcal{E}_{k}.
\end{equation}
Consider the quadratic trust function
\begin{equation}\label{eq:costfunc}
    f_{k}(x) = \lVert x -\bar{x}_{k} \rVert _{2},
\end{equation}
where $\bar{x}_{k}$ is a vector in $\mathcal{X}$ corresponding to optimal trust for individual $k$. 
We may estimate this function by considering the preference set as analogous to a set of affine classifications.
In feature vector space, this set corresponds to a set of hyperplanes separating the pairs of behavior points with maximal distance to each point. 

Given the $i^{th}$ preference pair $(x_{i}^{1}, x_{i}^{2})\in \mathcal{X}\times \mathcal{X}$, (\ref{eq:costComparison}) is equivalent to the halfspace
\begin{equation}\label{eq:hyperplane}
    g_{i}(x) = a_{i}^{T}x-b_{i} \leq 0 \;,
\end{equation}
where $a_{i} = x_{i}^{2} - x_{i}^{1}$ and $b_{i} = a_{i}^{T}(x_{i}^{1} + x_{i}^{2})/2$.
The closed halfspace indicates that $g_i(x)\leq 0$ (the region of the feature space containing preferred behaviors) is convex but not affine.

\subsection{Preference Polytope}
The $k^{th}$ individual's preference set can be described by a set of halfspaces (\ref{eq:hyperplane}) in feature vector space, with the $i^{th}$ hyperspace associated with the preference $(v^{1}_{i}, v^{2}_{i})\in \mathcal{E}_{k}$. Each preference instance reduces the halfspace of $\mathcal{X}$, further constraining $x$. The intersection of the closed halfspaces  define a \textit{preference polytope}, a region of feature vector space associated with greatest preference.
The intersection satisfies the system of linear inequalities created by the preferences, and can be represented by the polytope
\begin{equation} 
    \label{eq:poly}
    P_k = \left\{x\in \mathcal{X}|a_{i}^{T}x\leq b_i, \forall (v^{1}_i,v^{2}_i)\in \mathcal{E}_{k} \right\}.
\end{equation}
An example the intersection of eight preference's corresponding halfspaces is given by the shaded interior region in Figure \ref{fig:Chebyshev}.
The preference polytope can be unbounded for small  $|\mathcal{E}_{k}|$ and poorly distributed preference pairs. The preference polytope can also be empty for cyclic preference graphs and poorly selected embeddings.

The closed region of the polytope $P_{k}$ can be used to determine preferred swarm behaviors; this process is often termed preference learning \cite{herbrich1998supervised}. The polytope $P_{k}$ can be built iteratively, with new pairwise comparisons presented to the individual over time. Given the polytope $P_{k}(t)$ at sample time $t$, the addition of the $i^{th}$ preference at time $t+1$ forms the new polytope $P_{k}(t+1)= P_{k}(t)\bigcap \{x|g_i(x)\leq 0\}$. The strategic presentation of pairwise comparisons to the participant can rapidly reduce the volume of $P_{k}(t)$ over time. 

\subsection{Finding the Chebyshev Center}\label{subsec:Chebyshev}
\begin{figure}[!t]
\centering
\includegraphics[width=0.65\columnwidth]{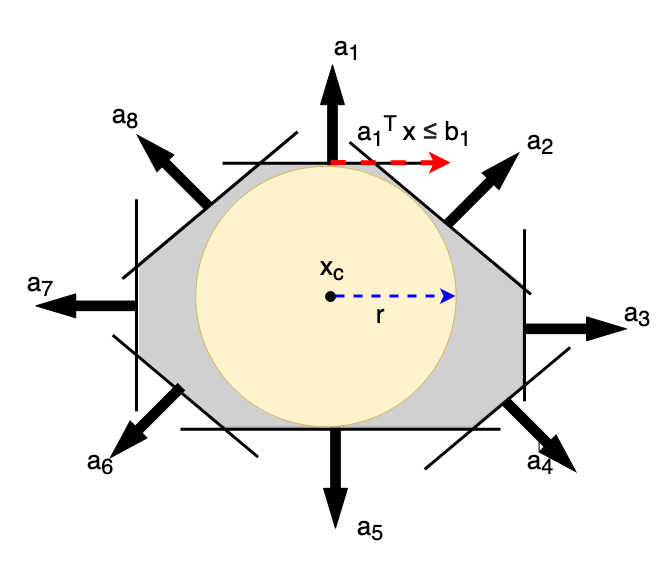}
\caption{The Chebyshev center $x_{c}$ with radius $r$ of the preference polytope $P$.} 
\label{fig:Chebyshev}
\end{figure}

As we have insufficient information within a bounded $P_k$ to find $\bar{x}_k$ in (\ref{eq:costfunc}), we may substitute an alternative point in $\mathcal{X}$.
We believe that the Chebyshev center of $P_{k}$ is a suitable candidate \cite{kingston_comparing_2009}.
The Chebyshev center of $P_k$ is the center of the largest inscribed ball in $P_k$, also referred to as the in-center point. A visual interpretation of this is depicted in Figure \ref{fig:Chebyshev}. 
Let the Chebyshev center $x_{c}$ lie at the center of the largest possible ball $\mathcal{B} = \{x_{c}+u\,|\,\lVert u \rVert_{2} \leq r\}$ inside $P_{k}$.
We may obtain $\mathcal{B}$ by maximising $r$.
For a weaker constraint, let $\mathcal{B}$ lie in the half-space
\begin{equation}
    \lVert u \rVert_{2} \leq r \implies a_{i}^{T}(x_{c} + u) \leq b_{i}.
\end{equation}
The corresponding largest possible ball is given by
$\text{sup}\{a_{i}^{T}u| \lVert u \rVert_{2}\leq r\} = r\lVert a_{i}\rVert_{2}.
$
Hence the Chebyshev center lies within the halfspace $\mathcal{B}\subset P_k$ if and only if $
    a_{i}^{T} x_{\text{c}} + r\lVert a_{i} \rVert_{2} \leq b_{i}, $ for all $(v_i^1,v_i^2)\in\mathcal{E}_k$.
Given the ball radius $r\geq 0$, $x_{\text{c}}$ can be found by solving the optimization
\begin{align}
    \left({x}_c,\overline{r}\right) &= \arg \max_{x,r} \;r, \\
    \textrm{s.t.} \: \: &a_{i}^{T} x +r \lVert a_{i} \rVert _{2} \leq b_{i}, \forall (v_i^1,v_i^2)\in\mathcal{E}_k \nonumber.
\end{align}
The optimization is a linear program with many algorithms that can reliably and efficiently solve the problem \cite{boyd2004convex}.
The resulting $x_c$ can then be used as a proxy to compare individuals' trust.

\section{Trust for Groups}\label{trust_gr}
We may analyze the trust preferences of a group of individuals. 
We first examine the distinctiveness of individuals' preferences using labeled pairwise preference data and consider the effect of self-contradictory responses.
Then considering unlabeled pairwise preferences, we observe properties of the entire group's preferences. 
Assuming that the individual’s point of maximal trust is drawn from a normal distribution, we use this to evaluate the cohesion of the group.

\subsection{Labeled Individual Preferences}
Given multiple individuals' respective preferences, we may devise a measure of the individuals' \textit{distinctiveness}.
The optimal trust value for each individual $k$ is denoted as $x+z^{k}$
where $x$ is assumed to be a global reference trust measure and $z^{k}$
the perturbation of individual $k$ away from this reference. 
If the magnitude of $z^{k}$ were small for all of the team, the group would exhibit similar trust behaviors. 
Individual with large $z^{k}$ would, in turn, have distinctive trust behaviors compared to the group as a whole. 
Examining the preferences across each individual $k$, the selection problem for their $i^{th}$
preference selection will generate the hyperspace $\left(a_{i}^{k}\right)^{T}\left(x+z^{k}\right)\leq b_{i}^{k}$.
An optimization problem can then be posed to find $x$
with small $\left\Vert z^{k}\right\Vert_1$ across all members using
\begin{align}
\left(\overline{x},\overline{z}\right) & =\text{argmin}_{x,z=[z^{1},\dots,z^{n}]}\sum_{k=1}^{n}\left\Vert z^{k}\right\Vert_1\label{eq:DistinctiveOptimization},\\
\text{s.t.\,} & \left(a_{i}^{k}\right)^{T}\left(x+z^{k}\right)\leq b_{i}^{k}, \forall (v_i^1,v_i^2)\in\mathcal{E}_k \nonumber .
\end{align}
The accumulative 1-norm is used here to minimize $z$ due to its sparsifying properties as it promotes sparse solutions with small or even zero $\left\Vert z^{k}\right\Vert_1$ \cite{boyd2004convex}. 
When $\left\Vert z^{k}\right\Vert_1=0$, the reference trust measure $x$ will satisfy all
of the selection preferences for individual $k$. This subset of individuals will share a non-trivial intersection of their trust polytopes and an additional Chebyshev center selection could be
performed to select a trust reference $x_c$ with the characteristics
described in $\S$\ref{subsec:Chebyshev}.

\subsection{Unlabeled Population Preferences}

We now examine the case where a preference may be expressed multiple times with contradictory responses. 
Consider the selection of the perturbation $z$ for each individual from a normal distribution
$Z\sim\mathcal{N}(0,\Sigma)$, where $\Sigma$ is a symmetric positive
definite matrix. From (\ref{eq:hyperplane}), 
the $i^{th}$ selection problem can subsequently be
posed as a preferential selection of one choice over the other when
the random variable 
$
X_{i}=g_{i}(x+z)=a_{i}^{T}\left(x+z\right)-b_{i}
$
is non-positive; the corresponding distribution is $\mathcal{N}(\mu=a_{i}^{T}x-b_{i},\sigma^{2}=a_{i}^{T}\Sigma a_{i})$.
By applying the inverse distribution mapping (\ref{eq:Inverse distribution mapping}),
the probability of a positive preference selection $p_{i}=p(X_{i}\leq0)$
is then 
$
-\Phi^{-1}(p_{i})\sigma=\mu.
$
Assuming that the covariance of $Z$ is bounded as $0\preceq\Sigma\preceq\alpha^{2}I$,
then $\sigma\leq\alpha\left\Vert a_{i}\right\Vert _{2}$. For $p_{i}\geq 0.5$ 
then $\Phi^{-1}(p_{i})\geq0$ and 
\begin{equation}
-\alpha\left\Vert a_{i}\right\Vert _{2}\Phi^{-1}(p_{i})\leq a_{i}^{T}x-b_{i}\leq0.\label{eq:Prob Constraint 1}
\end{equation}
Similarly, for $p_{i}<0.5$ then 
\begin{equation}
0\leq a_{i}^{T}x-b_{i}\leq-\alpha\left\Vert a_{i}\right\Vert _{2}\Phi^{-1}(p_{i}),\label{eq:Prob Constraint 2}
\end{equation}
such that each preference constrains $x$ to lie in what can be interpreted geometrically as a slab, i.e., a set of the form $\{x\in \mathbb{R}^{q}|\alpha\leq a^{T}x\leq\beta\}$ for scalars $\alpha\leq\beta$.
With the data sampled from a finite population, the probability $p_{i}$
is calculated based on a confidence interval $\left[p_{i}-\delta,p_{i}+\delta\right]$
projected onto the unit interval $\left[0,1\right]$ as $\Delta=\left[\max(0,p_{i}-\delta),\min(1,p_{i}+\delta)\right])$.
Here, $2\delta$ is the width of the confidence interval band and
is based on the margin of error calculated from a number of samples.
Applying the central limit theorem for the binomial distribution is one approach to calculate the width with $\delta=Z\sqrt{1/4n_{s}}$ where $Z$
is the $Z$-score associated with a confidence interval and $n_{s}$
is the number of samples of the $i^{th}$ preference \cite{brown_interval_2001}.

\begin{figure}[!t]
\centering
\includegraphics[width=0.75\columnwidth]{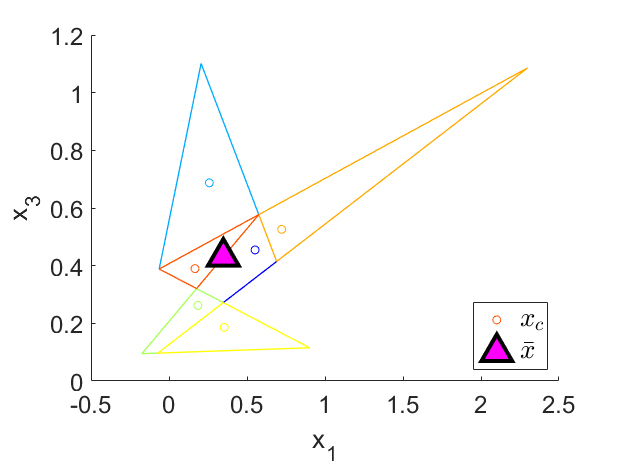}
\caption{Participant Chebyshev centers $x_c$ compared with the aggregated trust optimum $\bar{x}$.}
\label{fig:chebagg}
\end{figure}

We assume that $g_{i}(x)$ is constructed so that $X_i\leq0$ for most of the population (i.e., $p_{i}\geq 0.5 $), for example as per $\S$\ref{Preference graph} for $\bar{\mathcal{G}}=(\bar{V},\bar{\mathcal{E}},W)$ with $p_{i}=w_{i}\in W$. Using the constraints (\ref{eq:Prob Constraint 1}) and
(\ref{eq:Prob Constraint 2}) over $\Delta$, we may find the
average trust measure $\overline{x}$ and minimum covariance bound
$\overline{\alpha}$ for the population as  
\begin{align}
\label{eq:populationGraphOptimization}
\left(\overline{x},\overline{\alpha}\right) & = \arg \min_{x,\alpha}\;\alpha\\\nonumber
\text{s.t.\,\,} & a_{i}^{T}x-\alpha\left\Vert a_{i}\right\Vert _{2}\max(0,-\Phi^{-1}(p_{i}-\delta))\leq b_{i},\\\nonumber
 & a_{i}^{T}x+\alpha\left\Vert a_{i}\right\Vert _{2}\Phi^{-1}(\min(1,p_{i}+\delta))\geq b_{i},\\\nonumber
 & \hphantom{a_{i}^{T}x+\alpha\left\Vert a_{i}\right\Vert _{2}\Phi^{-1}}\forall(v_{i}^{1},v_{i}^{2})\in\bar{\mathcal{E}},p_{i}=w_{i}\in W.\nonumber
\end{align}
For the $i^{th}$ preference, the (positive) upper bound on the covariance $\alpha$ constrains the width of the slab containing $x$. Similarly, the closer $p_{i}$ is to $0.5$ 
(i.e. a split decision on the $i^{th}$ preference among individuals), the narrower the slab. 

\section{User Study}\label{study}

We have conducted a user study to observe human trust preferences regarding swarm behaviors. In this section we outline our procedure and summarize collected results.

\begin{figure}[!t]
\centering
\includegraphics[width=0.8\columnwidth]{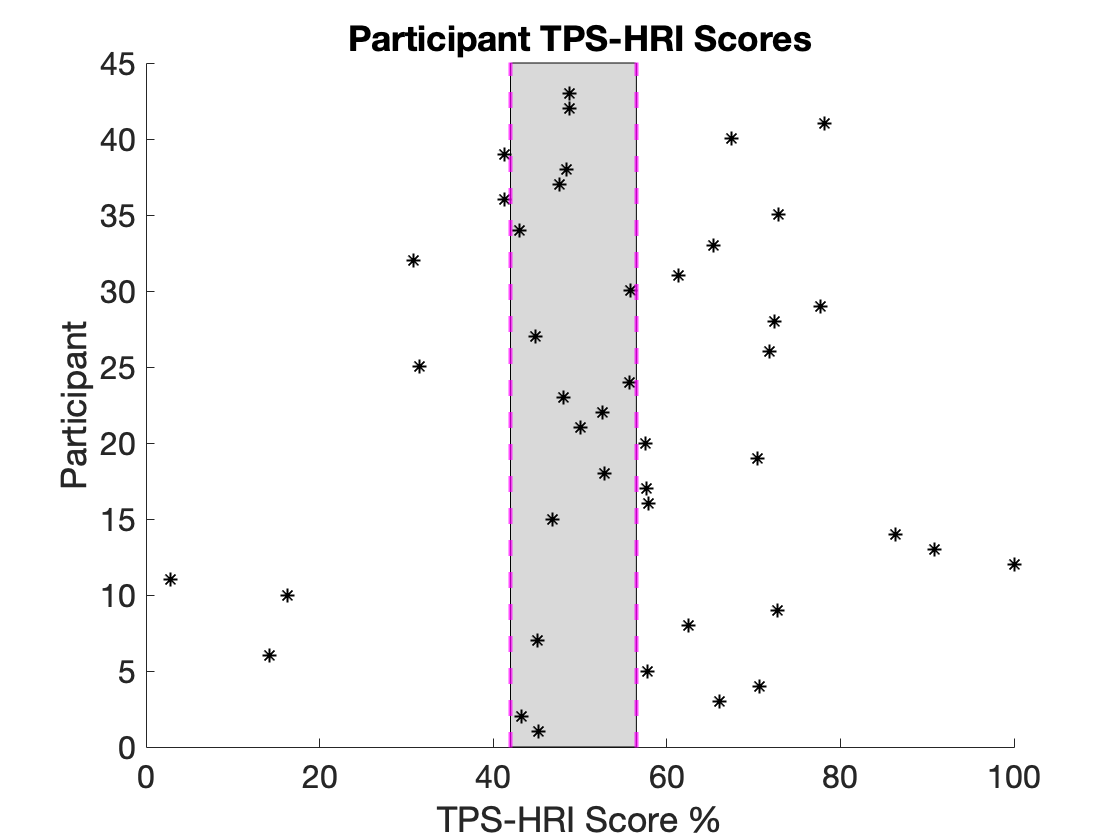}
\caption{Trust scores for participants as measured using TPS-HRI; the shaded region contains participants with distinctiveness $\left\Vert z^{k}\right\Vert_2 \leq 0.035$.} 
\label{fig:TrustScores}
\end{figure}

\begin{figure}[!t]
\centering
\includegraphics[width=0.8\columnwidth]{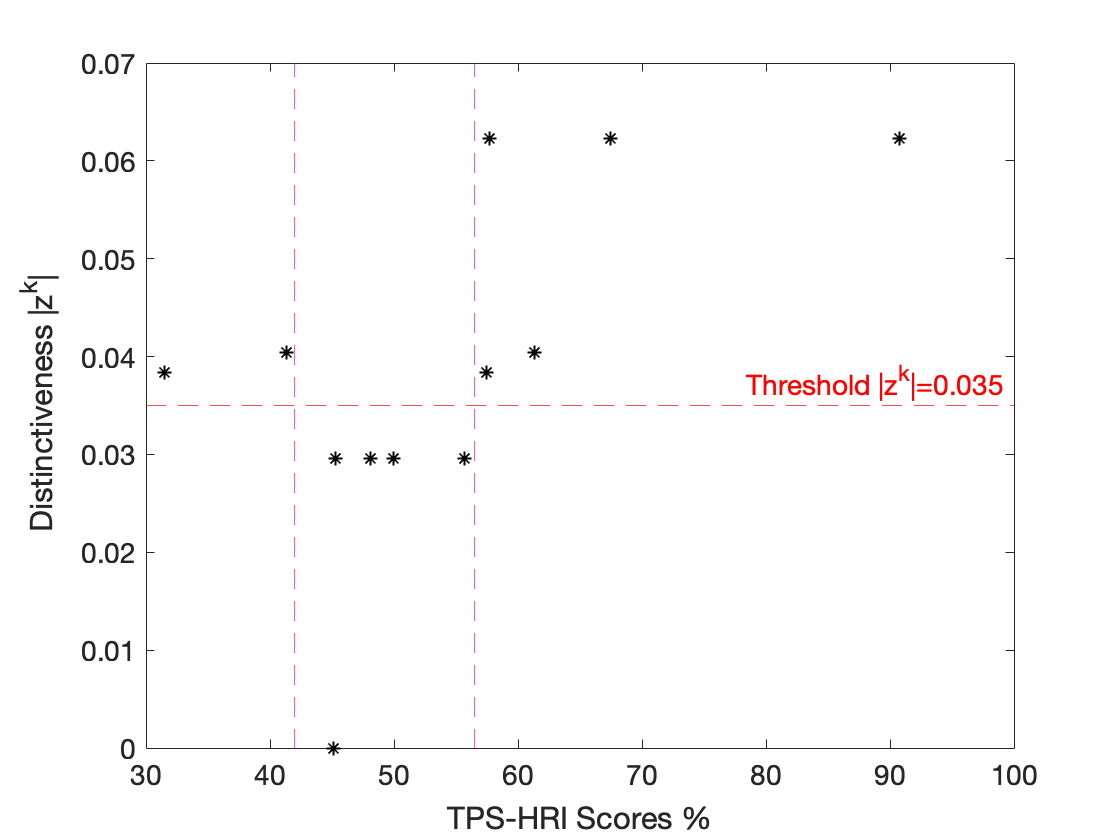}
\caption{Participant distinctiveness.} 
\label{fig:TPS_distinct}
\end{figure}

\subsection{Procedure}

We pursued an online survey methodology involving filmed videos of the swarm behaviors from $\S$\ref{swarmbehaviors}. 
All participants were over 18 years of age and no demographic information was requested. Forty-three participants responded to the survey, and were not remunerated or otherwise rewarded for survey completion.
In the first part, we presented the participant with a video of a swarm executing a leader following behavior (see  $\S$\ref{swarmbehaviors}). The fourteen questions relating to trust in the swarm are a modified version of the \textit{Trust Perception Scale-HRI} (TPS-HRI) \cite{schaefer_measuring_2016}, substituting the word \textit{`swarm'} for \textit{`robot'}.
In the second part, we presented the participant with pairs of swarm behavior videos and collected the participant's pairwise trust preferences, asking \textit{`Comparing the two swarms, which do you trust \textit{more}?'}. To avoid priming the participants we did not specify the notion of trust further.
The first six pairs of videos covered each combination among the first four of the five behaviors, shown to each participant in the same order.
Each participant could repeat the process with four more video pairs, but this was optional to avoid survey fatigue impacting response quality.

\paragraph*{Feature Vector Extraction} 
The feature vectors extracted by \textsc{Valma} encode sequential dependencies between successive video frames: for a training set of 20 videos, the percentage of dimensions differing between the \textit{original video's} feature vector and that of the \textit{same video played in reverse} lies in the range $[23.1\%,49.5\%]$. Since the only difference is in the order of video frames presented to \textsc{Valma}, and given that this has yielded distinct feature vectors, we infer that \textsc{Valma} can map distinct videos to distinct feature vectors.

\subsection{Results}
\begin{figure}[!t]
    \centering
    \subfigure[{}]{
        \includegraphics[width=0.32\columnwidth]{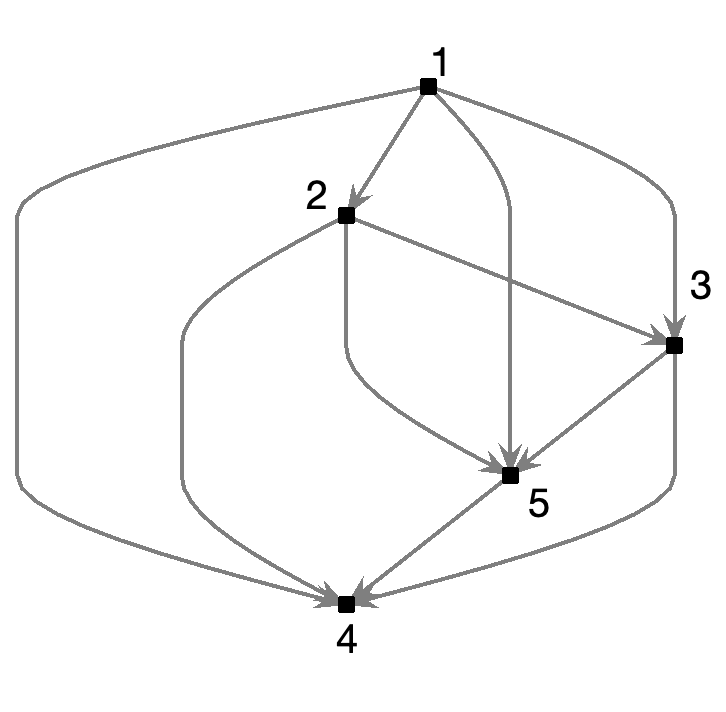}
        \label{fig:indG}
        }
    \subfigure[{}]{
        \includegraphics[width=0.4\columnwidth]{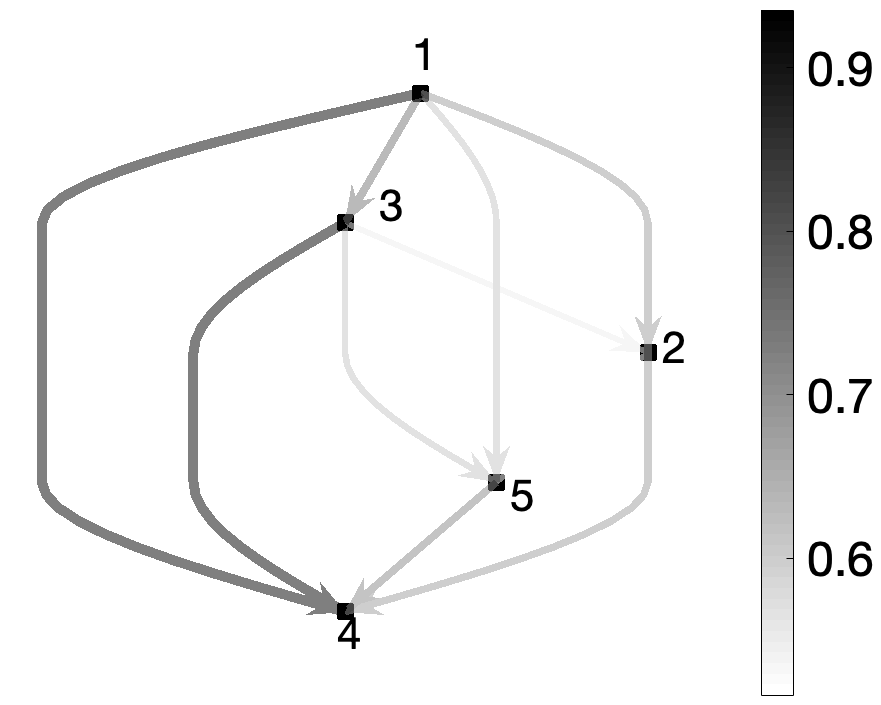}
        \label{fig:popG}
    }
    \caption{a) A participant's acyclic preference graph; 
    b) Weighted population preference graph (the gradient bar indicates the preference likelihood for the population).}
    \label{fig:PrefGraphs}
\end{figure}

We proceed to analyze data gathered from the online survey, focusing on the distinctiveness and cohesion of the cohort's trust preferences.
For each participant we have created a preference graph (exemplified by Figure \ref{fig:indG}) to determine the respective preference polytope.
For the subset of participants with bounded preference polytopes, their individual Chebyshev centres could be found.
In Figure \ref{fig:chebagg} the aggregated population trust optimum $\bar{x}$ from (\ref{eq:DistinctiveOptimization}) is compared with the Chebyshev centers for the preference polytopes belonging to a subset of participants.\footnote{For visualization purposes only two dimensions of $\mathcal{X}$ are depicted.}
For the same subset of participants, we also compare their distinctiveness $\left\Vert z{^k}\right\Vert_1$ with corresponding TPS-HRI trust scores in Figure \ref{fig:TPS_distinct}.
We observe that participants with distinctiveness $\left\Vert z^{k}\right\Vert_1 \leq 0.035$ and a trust score in the range $[42\%,56.5\%]$ express preferences compatible with the population's preferences. In contrast, participants with distinctiveness $\left\Vert z^{k}\right\Vert_1 > 0.035$ have preferences differing from the population.
In Figure \ref{fig:TrustScores} we extrapolate the notion of distinctiveness and trust bounds to the whole population.
Participants with trust values in the range $[42\%,56.5\%]$ are associated with low distinctiveness from the population's preferences in Figure \ref{fig:TPS_distinct}, hence have similar trust preferences to an average participant in the population. In this way low distinctiveness can become a criterion for selecting teams of participants.



In Figure \ref{fig:popG} a partial ordering over the swarm behaviors is generated from an aggregation of unlabeled preferences from all participants, and represented as a preference graph $\bar{\mathcal{G}}=(\bar{V},\bar{\mathcal{E}},W)$ as depicted.
We may then use the edge-weighted population preference graph to derive an average trust measure $\mu =\bar{x}$ and minimum covariance bound $\sigma\leq\bar{\alpha}$ 
from (\ref{eq:populationGraphOptimization}).
In Table \ref{tab:repeatedPrefs1} we compare $\mu$ with the aggregated Chebyshev center $\bar{x}_{c}$ of the unweighted preference graph $(\bar{V},\bar{\mathcal{E}})$. 
We observe in Table \ref{tab:repeatedPrefs2} that $54.1\%$ of participant's individual Chebyshev centers lie within the upper bound of one standard deviation $\alpha$ of the mean and $100\%$ lie within $2\alpha$. This matches well the theoretical bounds $p(-s<\left\Vert X-\mu \right\Vert_2/\sigma<s)=\Phi(s)-\Phi(-s)$ for $s\in\{1,2\}$, hence the aggregated unlabeled preference data is consistent with the model presented in Figure \ref{fig:popG}. 
The relatively small distance between the mean and the population's Chebyshev center, $\left\Vert \mu-\bar{x}_{c}\right\Vert_2 \leq0.1\bar{\alpha}$, shows that the population preference graph $\bar{\mathcal{G}}$ and optimal solution of (\ref{eq:populationGraphOptimization}) is representative of the true value of $\bar{x}_k$. This suggests that we may analyze preference similarity to evaluate a population's cohesiveness.

\begin{table}[!t]
\centering
\caption{Population Trust Statistical Information}
\label{tab:repeatedPrefs1}
\begin{tabular}{|c|c|c|}
\hline
\multirow{1}{*}{\textbf{Aggregate} $\mathbf{\bar{x}_c}$}
 & \textbf{Mean,}\,$\mu$ & \textbf{Covariance Bound,}\,$\bar{\alpha}$\\ \hline
(-0.4076,  0.1697)  & (-0.4383, 0.1788) & 0.3406\\\hline
\end{tabular}
\end{table}

\begin{table}[!t]
\centering
\caption {Population Trust Preference Distribution}
\label{tab:repeatedPrefs2}
\begin{tabular}{|c|c|c|}
\hline
\textbf{s} & $\mathbf{p(-s<\left\Vert x_c-\mu\right\Vert_2/\bar{\alpha}<s)}$ & $\Phi(s)-\Phi(-s)$\\\hline
1 & 0.5405 & 0.6812  \\\hline
2 & 1.0000 & 0.9545  \\\hline
\end{tabular}
\end{table}

\section{Conclusions}\label{conclusions}

In this work, we have studied a model of human trust in a robotic swarm using preferences. 
Generating a unique feature vector for each swarm behavior using \textsc{Valma}, we have embedded the swarm behaviors into a feature space and have formulated a polytope model with a Chebyshev center. 
Extending our consideration to groups of individuals, we have formulated a new \textit{distinctiveness} metric to measure individuals' labeled pairwise trust preferences with respect to a wider population. 
Aggregating all pairwise trust preferences for a group, we have posed a sparse optimization problem informed by the population's weighted preference graph.
This yields an average trust measure and minimum covariance bound, enabling analysis of the group's \textit{cohesion}. 
Results from our user study suggest that individuals with similar trust profiles may be grouped by low distinctiveness.

We anticipate three main areas for future work: measuring steady-state trust preferences in longer-duration interactions, 
modeling a population's aggregated trust preferences using explicitly distinct clusters,
and identifying an ideal truncation length for feature vectors produced by \textsc{Valma}.

\addtolength{\textheight}{-12cm}   

\bibliographystyle{IEEEtran}
\bibliography{root}
\end{document}